\documentclass[10pt,conference ]{IEEEtran}
\usepackage{amsfonts}
\usepackage{times}
\usepackage{graphicx}
\usepackage{latexsym}
\usepackage{dsfont}
\usepackage{amssymb}
\usepackage{amsmath}
\usepackage{cite}
\usepackage{verbatim}
\usepackage{subfigure}

\newcommand{\figref}[1]{{Fig.}~\ref{#1}}

\def\bb0{{\mathbb{0}}}

\def\ba{{\mathbf{a}}}
\def\bb{{\mathbf{b}}}

\def\bp{{\mathbf{p}}}

\def\b0{{\mathbf{0}}}

\def\bS{{\mathbf{S}}}

\def\bX{{\mathbf{X}}}

\def\sf0{{\mathsf{0}}}

\usepackage{multirow}
\usepackage{multicol}
\usepackage{booktabs}
\usepackage{epstopdf}
\usepackage{enumerate}
\usepackage{algorithm}
\usepackage{amsmath}
\usepackage{float}
\usepackage{color}
\usepackage{makeidx}
\usepackage{bm}
\usepackage{cleveref}
\usepackage{url}
\usepackage{steinmetz}
\usepackage{varwidth}
\usepackage{soul}
\graphicspath{{figs/}}
\usepackage{tabularx}
\usepackage{gensymb}
\usepackage{balance}

\newcommand{\comm}[1]{}

\begin{document}

\title{Vehicle Cameras Guide mmWave Beams: \\Approach and Real-World V2V Demonstration}
\author{ Tawfik Osman, Gouranga Charan, and Ahmed Alkhateeb\\ Arizona State University, Emails: \{tmosman, gcharan, alkhateeb\}@asu.edu}

\maketitle

\begin{abstract}
	Accurately aligning millimeter-wave (mmWave) and terahertz (THz) narrow beams is essential to satisfy reliability and high data rates of 5G and beyond wireless communication systems. However, achieving this objective is difficult, especially in vehicle-to-vehicle (V2V) communication scenarios, where both transmitter and receiver are constantly mobile. Recently, additional sensing modalities, such as visual sensors, have attracted significant interest due to their capability to provide accurate information about the wireless environment. To that end, in this paper, we develop a deep learning solution for V2V scenarios to predict future beams using images from a 360 camera attached to the vehicle. The developed solution is evaluated on a real-world multi-modal mmWave V2V communication dataset comprising co-existing 360 camera and mmWave beam training data. The proposed vision-aided solution achieves $\approx 85\%$ top-5 beam prediction accuracy while significantly reducing the beam training overhead. This highlights the potential of utilizing vision for enabling highly-mobile V2V communications.
	
\end{abstract}


\section{Introduction} \label{sec:Intro}

Millimeter wave (mmWave) and terahertz (THz) communications adopt large antenna arrays and use narrow directive beams to guarantee sufficient receive power \cite{Rappaport2019}. Accurately aligning the narrow beams is crucial for achieving high data rates in highly-mobile applications such as vehicle-to-vehicle (V2V) communication scenarios. However, selecting the optimal beams for these systems with large antenna arrays is typically associated with a large training overhead, making it challenging for mmWave/THz communication systems to support these future applications. Prior work on reducing the mmWave beam training overhead have typically focused on constructing adaptive beam codebooks \cite{Zhang2021a}, designing beam tracking techniques \cite{Jayaprakasam2017}, and leveraging the channel sparsity and efficient compressive sensing tools \cite{Alkhateeb2014,HeathJr2016}. However, these classical approaches typically can only reduce the training overhead by one order of magnitude, which might not be sufficient for systems with large antenna arrays and highly mobile applications. This motivates exploring new approaches to overcome this beam training overhead and enable highly-mobile mmWave/THz V2V communication systems.

\begin{figure}[!t]
	\centering
	\includegraphics[width=1.0\linewidth]{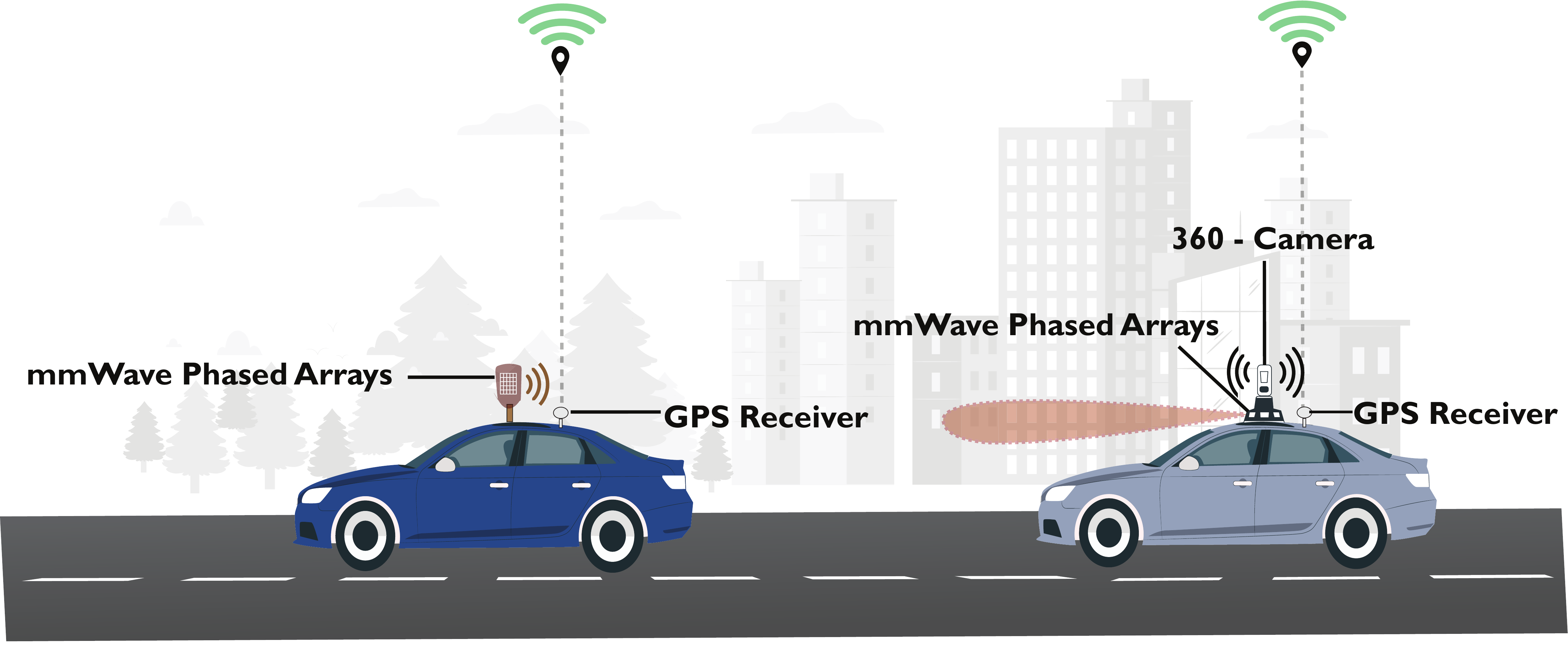}
	\caption{An illustration of the adopted system model. The receiver vehicle utilizes the visual data to aid mmWave beam tracking. }
	\label{sys_fig}
\end{figure}

The use of machine learning (ML) techniques to address the beam prediction task has become increasingly popular in recent years \cite{Alkhateeb2018, RobertPos, Charan22, Jiang2022a, Demirhan2022b}. These solutions primarily aim to leverage additional information to enhance the awareness of the wireless environment. Furthermore, the additional sensing modalities have been shown to be instrumental in the development of real-time digital twins (digital replicas) of the physical environments that can be utilized for making efficient communication and sensing decisions \cite{jiang2023digital, alkhateeb2023real}. In \cite{Alkhateeb2018}, the authors propose to utilize the receive wireless signature to predict the optimal beam indices at the basestation. Another approach is to utilize position information, as demonstrated in \cite{RobertPos}, to predict the optimal beam index. Although incorporating location data can help reduce the training overhead, relying solely on GPS data may lead to inaccurate predictions due to inherent errors. To eliminate the beam training overhead, some researchers leverage other sensing modalities such as cameras, LiDAR, and radar. For instance, \cite{Charan22} utilizes RGB images captured by a camera at the basestation to guide beam prediction, while \cite{Jiang2022a} uses LiDAR point cloud data to facilitate beam prediction and tracking. Furthermore, \cite{Demirhan2022b} proposes using radar installed at the basestation to predict the optimal beam index. However, all these solutions are designed for vehicle-to-infrastructure (V2I) scenarios and are limited to a single-user setting.

In this paper, we propose to leverage visual data captured by cameras installed on the vehicles to realize beam tracking in V2V communication scenarios. Given a sequence of recent RGB images of the wireless environment and the initial optimal receive power vector, the objective is to predict the optimal beam corresponding to the latest data capture. The main contributions of the paper can be summarized as follows:
\begin{itemize}
	\item Formulating the vision-aided V2V beam tracking problem in mmWave/THz wireless networks considering practical visual and communication models. 
	\item Given a sequence of image samples and the receive power vector corresponding to the first sample in the sequence, developing a machine learning-based solution that is capable of (i) detecting objects of interest in the wireless environment and extracting the relevant features, (ii) identifying the transmitter in the scene, and (iii) efficiently predicting the optimal beam for the future sample.  
	\item Providing the first real-world evaluation of vision-aided V2V beam tracking based on our large-scale dataset, DeepSense 6G \cite{DeepSense}, that consists of co-existing multi-modal sensing and wireless communication data. 
\end{itemize}
Based on the adopted real-world dataset, the developed solution achieves $98$\% transmitter identification accuracy. Further, the proposed solution achieves a top-5 prediction accuracy of $\approx$ 85\% while reducing the beam training overhead. This highlights the capability of the proposed sensing-aided beam tracking approaches to reduce the beam training overhead significantly.

\begin{figure*}[!t]
	\centering
	\includegraphics[width=0.9\linewidth]{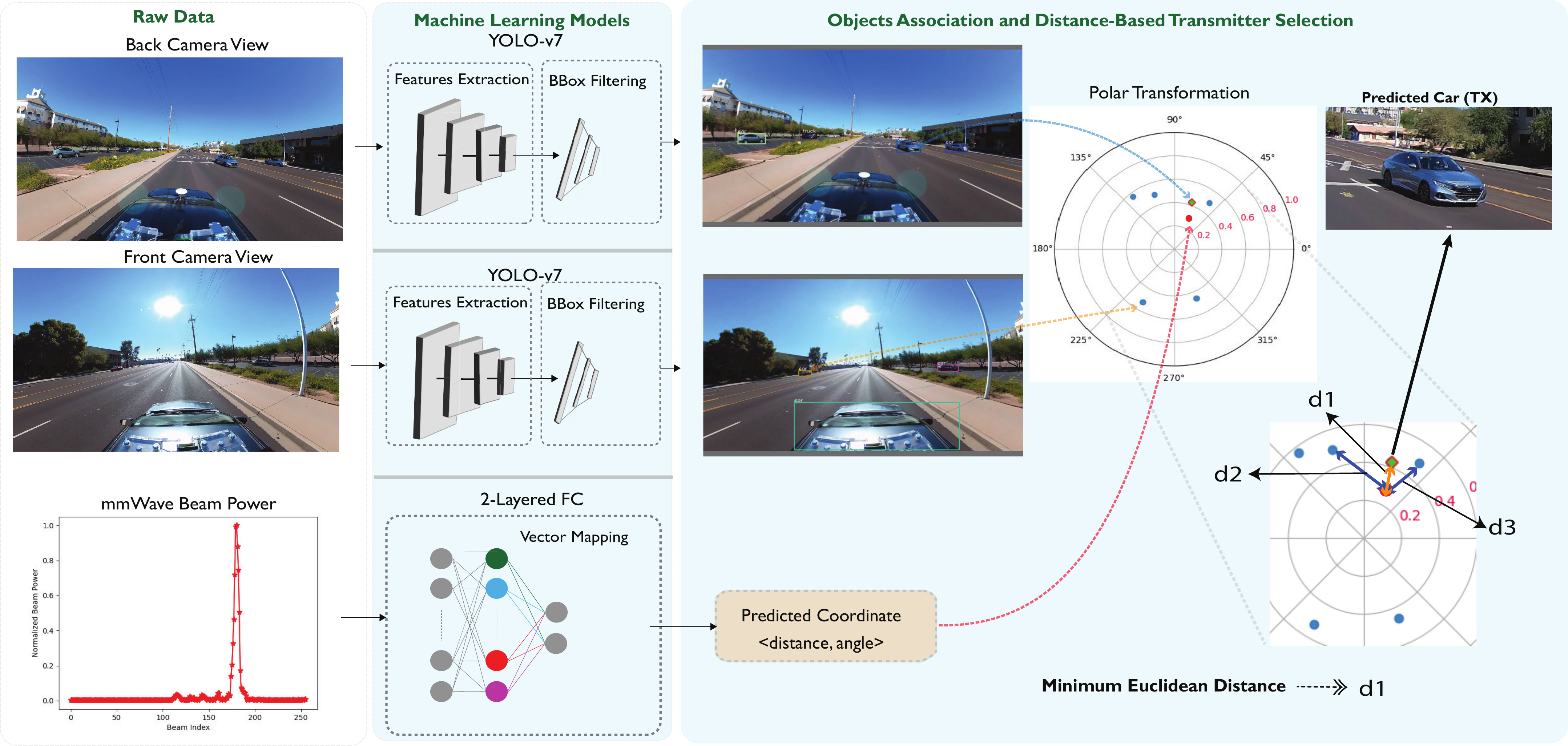}
	\caption{This figure illustrates the stages of the transmitter identification task. The front and back image is fed into a deep learning model to extract the bounding box coordinates. The mmWave Beam power vector is provided as an input to a fully connected neural network to predict the transmitter coordinates, subsequently the candidate transmitter's coordinate  is selected from the possible coordinates using the nearest distance-based algorithm.}
	\label{fig:txfig}
\end{figure*}

\section{System Model and Problem Formulation}\label{sec:sys_ch_mod}
This work considers a V2V communication scenario consisting of a vehicle acting as the transmitter and another vehicle acting as the receiver in a real wireless communication environment as shown in \figref{sys_fig}. In particular, the communication system model consists of: (i) A mobile mmWave receiver equipped with a set of $M$-element Uniform Linear Array (ULA), each directed towards a different direction to provide $360\degree$ coverage, a $360\degree$ RGB camera, and an RTK GPS receiver. (ii) A mobile transmitter equipped with an omni-directional antenna. In this section, we first present the adopted wireless communication model. Then, we formulate the sensing-aided V2V beam tracking problem. 

\subsection{Communication model} \label{sec:system_model}
The system model adopted in this paper constitutes a receiver vehicle that employs four mmWave transceivers, each with $M$ antennas to communicate with the transmitting vehicle, equipped with a single quasi-omnidirectional mmWave transceiver. Let $d \in \{front, back, left, right\}$ denote the communication and sensing direction of the receiver vehicle. Considering a geometric channel model, the channel $\mathbf h_{d} \in \mathbb C^{M\times 1}$ between the transmitter and receiver can be written as:
\begin{equation}
	\mathbf{h}_{d} = \sum_{\ell=1}^{L_{d}} \alpha_{\ell,d} \ba\left(\theta_{\ell,d}^{az}, \phi_{\ell,d}^{el} \right),
\end{equation} 
where $L_{d}$ is the number of channel paths, $\theta_{\ell,d}^{az}$, $\phi_{\ell,d}^{el}$ are the azimuth and the elevation angle of arrival, respectively, of the $\ell$th channel path.  $\alpha_{\ell,d}$ is the complex path gain. The communication system adopted in this work uses a predefined beamforming codebook $\boldsymbol{\mathcal F}=\{\mathbf f_m\}_{m=1}^{Q}$, where $\mathbf{f}_m \in \mathbb C^{M\times 1}$ and $Q$ is the total number of beamforming vectors in the codebook. It is important to note that the total number of beamforming vectors is $4Q$, with $Q$ beams in each direction. Now, given the geometric channel model $\mathbf{h}_{d}$, assume that the transmitter transmits the complex data symbol $x\in \mathbb C$ to the receiver, and the $d$-directional antenna array of the receiver receives the signal via the beamforming vector $\mathbf f_{d,m}$. The received signal $y_d$ can then be written as
\begin{equation}
	y_{d} = \mathbf f_{d,m}^H\mathbf h_{d}x + n_k,
\end{equation}
where $n_k$ is a noise sample drawn from a complex Gaussian distribution $\mathcal N_\mathbb C(0,\sigma^2)$. The transmitted complex symbol $x\in \mathbb C$ needs to satisfy the following constraint $\mathbb E\left[ |x|^2 \right] = P$, where $P$ is the average symbol power. The optimal beam index can be represented as the results of the maximization problem given as
\begin{equation}
	\underset{d, m}{\text{max}} \ | \mathbf{f}_{d,m}^H\mathbf h_{d} |^2,
	\label{eq:optimal_beam}
\end{equation}
where the solution to the maximization problem can be obtained by an exhaustive search over the beam codebook.

\subsection{Problem formulation} \label{sec:prob_form}
As presented in \eqref{eq:optimal_beam}, the optimal beam can be computed by either utilizing the explicit channel knowledge, which is generally hard to acquire or by performing an exhaustive beam search that results in large beam training overhead. In this work, instead of following the conventional beam training approach, we propose to utilize additional sensing modalities such as $360 \degree$ RGB images to predict the optimal beam index. In particular, at any given time $t$, we aim to predict the optimal beam index based on the current and previous visual data captured by the camera installed at the receiver vehicle. It is important to note here that the highly mobile nature of the V2V communication scenarios necessitates frequent updates to the communication beam after the initial connection has been established. These frequent updates will further increase the beam training overhead impacting the reliability and latency of the wireless communication systems. One promising solution to minimize this overhead is to track the transmitter vehicle over time using the RGB images and then predict the optimal beam, eliminating the need for any additional beam training. However, to track the transmitter vehicle over time, it is crucial first to identify the transmitter in the RGB images. For this, in addition to the RGB images, we propose to utilize the previous optimal receive power vector, which can be estimated when the connection between the vehicles is first established.

In this work, to facilitate the processing of the captured $360\degree$ image, we split it into two $180\degree$ field-of-view (FoV) images, covering the front and back sides, respectively. This allows us to focus on each side of the receiver vehicle independently and reduce the computational complexity of our approach. Let $\bX_{f}[t] \in \mathbb{R}^{W \times H \times C} \text{and} \ \bX_{b}[t] \in \mathbb{R}^{W \times H \times C} $ denote the front and back RGB images of the environment, respectively, captured at time instant $t$, where $W$, $H$, and $C$ are the width, height, and the number of color channels for the image. Further, let  $\bp[t] \in \mathbb{R}^{1 \times 4Q}$ denote the mmWave receive power vector at the receiver vehicle. At any time instant $\tau\in \mathbb Z$, the basestation captures a sequence of RGB images, and the mmWave receive power vector corresponding to the time instant of the first image capture, $\bS[\tau]$, defined as
\begin{equation}
	{\bS}[\tau] = \left\{ \left\{ \bX[t] \right\}_{t = \tau-r+1}^{\tau}, \bp[\tau-r+1] \right\}, 
\end{equation}
where $\bX[t] \in \{\bX_{f}[t], \bX_{b}[t]  \}$ and $r \in \mathbb Z$ is the length of the input sequence or the observation window to predict the optimal beam index. In particular, at any given time instant $\tau$, the goal in this work is find a mapping function $f_{\Theta}$ that utilizes the available sensory data samples $\bS[\tau]$ to predict (estimate) the future optimal beam index $ \hat{\mathbf f}[\tau + 1] \in \boldsymbol{\mathcal F}$ with high fidelity. The mapping function can be formally expressed as
\begin{equation}
	f_{\Theta}: \bS[\tau] \rightarrow  \hat{\mathbf f}[\tau + 1].
\end{equation}
In this work, we propose to utilize a machine learning-based model to learn this prediction function $f_{\Theta}$. The objective is to maximize the number of correct predictions over all the sample in $ \mathcal D = \left\lbrace \left (\bS_u, \mathbf f^{\ast}_u \right) \right\rbrace_{u=1}^U$, where $U$ is the total number of samples in the dataset. This can be formally written as 
\begin{equation}\label{eq:prob_form_1}
	f^{\star}_{\Theta^{\star}} = \underset{f_{\Theta}}{\text{argmax}}\\ \prod_{u=1}^U \mathbb P\left( \hat{\mathbf f}_u = \mathbf f^{\star}_u | \bS_u \right),
\end{equation}
where the joint probability in \eqref{eq:prob_form_1} is factored out to convey the identical and independent (i.i.d.) nature of the samples in dataset $\mathcal D$.

\begin{figure*}[!t]
	\centering
	\includegraphics[width=0.9\linewidth]{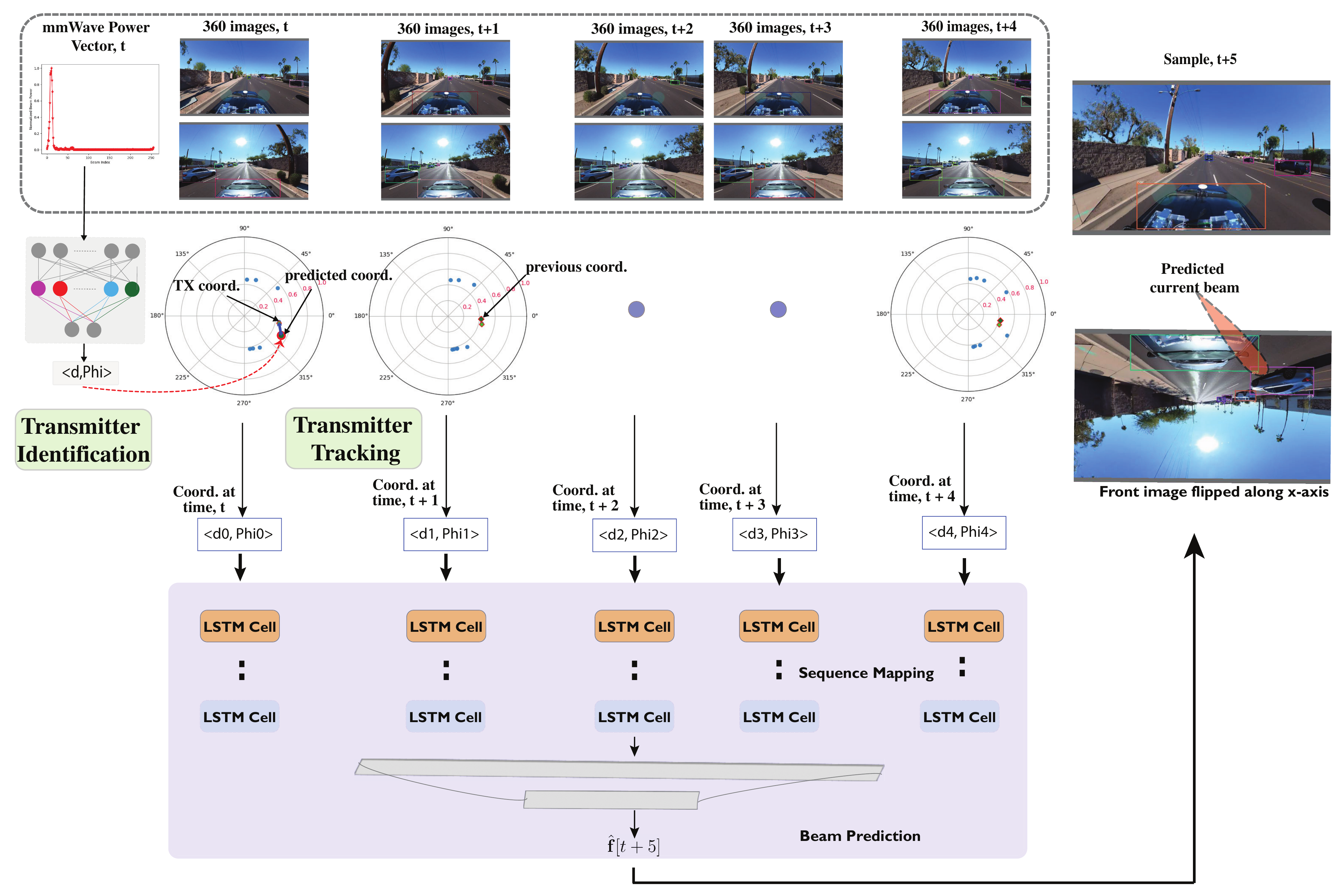}
	\caption{This figure presents an overview of the proposed deep learning and distance-based tracking solution. The bounding box coordinates for the front and back images are first transformed into a 2D polar plane. The mmWave beam power is fed into the transmitter identification model, to predict the current transmitter coordinates. The nearest distance-based algorithm is employed to track the transmitter coordinates in the 2D polar planes over the next four image samples. The tracked transmitter coordinates are fed to the LSTM and beam prediction sub-networks}
	\label{solution_fig}
\end{figure*}

\section{Vision-Aided V2V Beam Tracking: Proposed Solution} \label{sec:proposed_soln}
In this section, we propose a sensing-aided V2V beam tracking solution that operates through a sequence of four distinct sub-tasks. The first sub-task is an object detection solution, and it involves the detection of relevant objects of interest in the field of view (FoV) of the mmWave basestation situated on the receiver vehicle, using the visual data. The second sub-task is user identification, which involves utilizing the mmWave receive power vector to identify the transmitter within the wireless environment \cite{user_identification}. In the third sub-task, we utilize the coordinates predicted in the second task to track the object of the transmitter vehicle across the subsequent image samples. The final sub-task implements a beam prediction solution to predict the optimal beam index based on the sequence of bounding box center coordinates obtained from the preceding sub-tasks. \\

\textbf{Object Detection:}\label{object_model}
The visual data contains detailed information about the wireless environment, including vehicles that may house the transmitter unit. The first step in our solution involves identifying all objects belonging to the same visual class as the transmitter unit. To accomplish this, we employ a pre-trained object detection model called YOLOv7 \cite{yolov7}. The model has been adapted and fine-tuned to specifically detect relevant objects within the surrounding environment, including cars, trucks, and buses. We extract a 4-dimensional vector consisting of the bottom-left coordinates $[x_1, y_1]$ and the top-right coordinates $[x_2, y_2]$ for each object detected by the YOLOv7 model. These vectors are then processed to obtain the center coordinates of the objects in the 2D vector space, which are subsequently normalized to fall within the range of $[0, 1]$. The normalized center coordinates are further transformed into polar coordinates $(d,\theta)$, where $d$ is the radius and $\theta$ is the angle in degrees. This transformation is performed using the bottom-center coordinate as the reference point. Furthermore, we concatenated the polar coordinates of all the objects to form one dimensional vector, $\mathbf{d} \in \mathbb R^{2M\times 1}$, where $M$ is the number of objects detected. The length of the vector, $\mathbf{d}$ for each data sample, is determined by the count of candidate objects detected in the visual data. It is important to note here that this step is performed for all the $r$ image samples in the input sequence.

\textbf{Transmitter Identification:} \label{transmitter_model}
In the implementation of our single sample-based transmitter identification solution, we execute a two-step process. The first step involves utilizing a fully connected neural network (FCN) that is trained through supervised learning to estimate the probable polar coordinate of the transmission candidate using the normalized mmWave beam power vector. To facilitate the training and testing of the FCN model, we manually annotate $600$ image samples, generating a dataset that includes the ground-truth mmWave beam power and the corresponding transmitter's polar coordinates. The mmWave receive power vector and the transmitter coordinates are normalized using the global normalization and the min-max normalization, respectively. The specific training and testing parameters for the model are outlined in Table \ref{tab_nn_train_params}. Next, we employ a nearest distance-based algorithm to identify the potential transmitter coordinates in the polar transformation plane. To achieve this, we calculate the Euclidean distance between the predicted coordinates and all the potential coordinates in $\mathbf{d}$. The coordinate with the minimum Euclidean distance is then selected as the candidate transmitter coordinate ($d_{0}, \theta_{0}$) in the first image sample of the input sequence.

\textbf{Transmitter Tracking:} \label{track_model}
As outlined in Section~\ref{sec:prob_form}, the ground-truth mmWave receive power vector is accessible only for the first sample in the sequence. Consequently, the proposed user (transmitter) identification solution can only be performed for the first data sample in the sequence. Next, in order to identify the transmitter in the consecutive $r -1$ samples, we propose a transmitter-tracking solution. The proposed solution involves tracking the transmitter object over the next $r-1$ image samples, using the predicted polar coordinates from the transmitter identification model ($d_0, \theta_0$) and the vectors [ $\mathbf{d_1}, \ldots, \mathbf{d_{r-1}} $ ] obtained from the object detection task. The solution utilizes the nearest distance algorithm to track the polar coordinates of the potential transmitter over the next $r-1$ image samples, as shown in Fig.~\ref{solution_fig}. For this, the Euclidean distance between the polar coordinates of all objects in the current sample and the predicted coordinate of the transmitter in the last sample is computed. The object with the smallest distance to the previously identified transmitter coordinate is then selected as the coordinate of the transmitter for the current sample. The details on how this bounding box coordinates translation and tracking are depicted in Fig.~\ref{solution_fig}. It is important to note that we have converted the bounding box coordinates from both the front and back images into a single 2D polar plane. This conversion aids in tracking objects as they move between the front and back images. One challenge encountered involved instances where different sections of a single car were detected in both the front and back images. To address this, we devised an algorithm to coherently map the bounding box coordinates onto the 2D polar plane. This ensured that all relevant objects in the $360\degree$ image were represented as one point on the 2D plane.

\textbf{Recurrent Prediction:} In the final task, we leverage a recurrent neural network(RNN) to map the sequential transmitter coordinates tracked over a window of $r$ image samples in \ref{track_model} to predict the future optimal mmWave beam index. The recurrent neural network adopted in this work is the Long Short-Term network (LSTM) \cite{LSTM}. The input to the LSTM model is the sequence of $r$ polar coordinates of the transmitter, $\{[d_0, \phi_0]$,...,$[d_{r-1}, \phi_{r-1}]$\}, and the output is the future optimal beam index. We designed, trained, and tested a two-layered LSTM model using the parameters stated in Table \ref{tab_nn_train_params}.

\section{Testbed Description and Development Dataset}\label{sec:dataset}
To evaluate the effectiveness of the proposed sensing-aided beam tracking solution, we utilize the DeepSense 6G dataset \cite{DeepSense}. DeepSense 6G is the first large-scale real-world multi-modal dataset developed for sensing-aided wireless communication applications. It consists of co-existing multi-modal data such as mmWave wireless communication, GPS data, vision, Radar, and LiDAR collected in a real-world wireless environment. In this section, we describe the testbed and the development dataset. 

\begin{figure}[!t]
	\centering
	\includegraphics[width=1.0\linewidth]{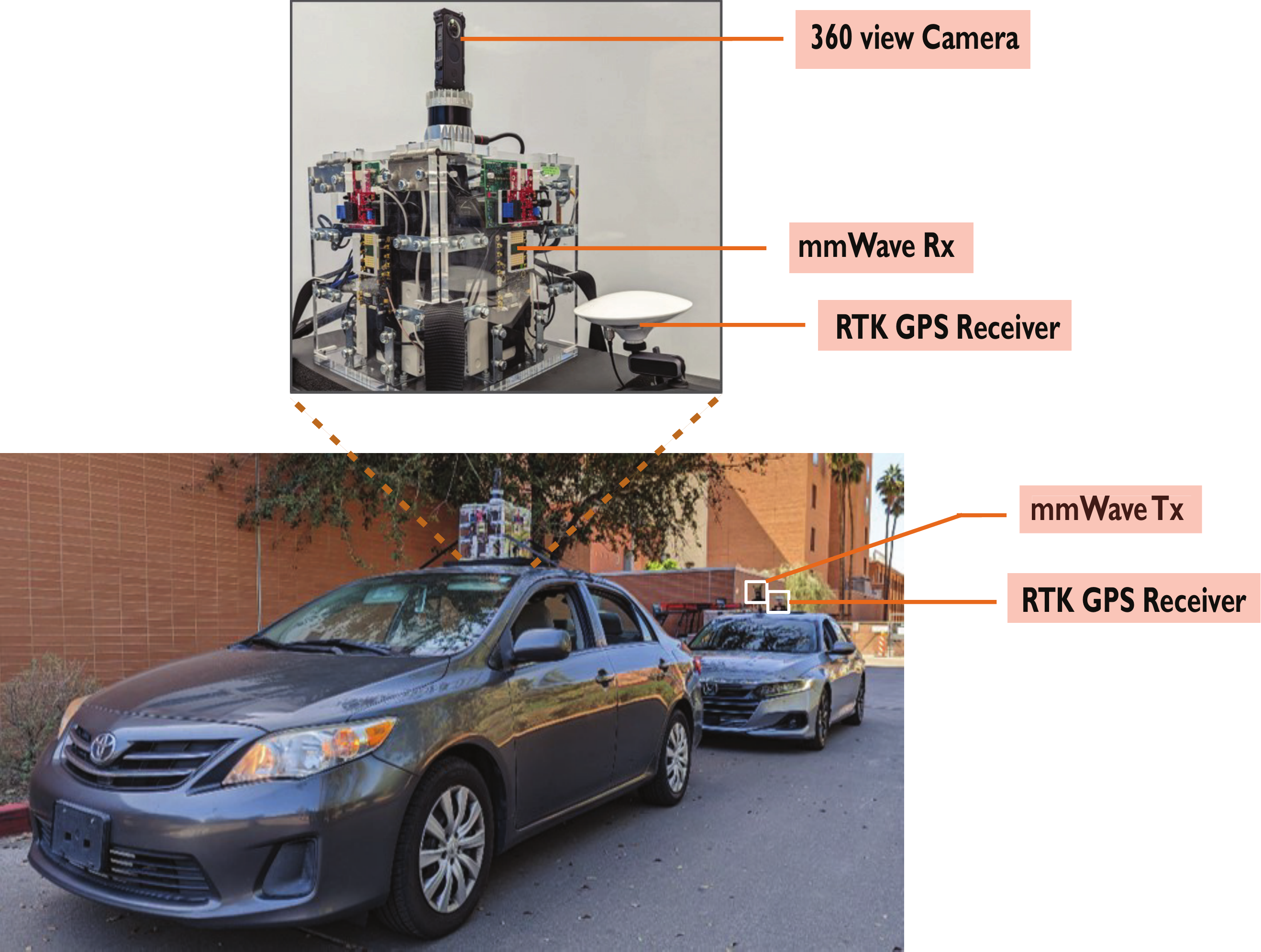}
	\caption{This figure shows a detailed description of the DeepSense 6G Testbed $6$ adopted to collect the real-world multi-modal V2V data samples. The front car (receiver unit) is equipped with four $60$ GHz mmWave Phased arrays, $360\degree$ RGB camera, four mmWave FMCW radars, one 3D LiDAR, and a GPS receiver. The back car (transmitter unit) is equipped with a $60$ GHz quasi-omni antennas and a GPS receiver. }
	\label{fig:testbed_fig}
\end{figure}

\textbf{DeepSense 6G Testbed 6:} This paper adopts scenario $36$ of the DeepSense 6G dataset designed specifically to study high-frequency V2V communication in the real world. The DeepSense testbed $6$, as shown in Fig.~\ref{fig:testbed_fig} is utilized in the data collections. It consists of two units: (i) Unit $1$, mobile receiver (vehicle) equipped with four $60$ GHz mmWave Phased arrays facing four different directions, i.e., front, back, left, right. Each phased array adopts a uniform linear array (ULA) with $M=16$ elements and utilizes an over-sampled pre-defined codebook of $Q=64$ beam directions to receive the transmitted signal. It is further equipped with a $360\degree$ RGB camera, four mmWave single-chip FMCW radars (operating between 76-81GHz), one 3D LiDAR with 32 vertical and 1024 azimuth channels, and one GPS RTK kit. (ii) Unit $2$, a mobile transmitter equipped with a $60$ GHz quasi-omni antennas always oriented towards the receiver unit and a GPS receiver to capture the real-time position information.

\textbf{Development Dataset:} The evaluation of the proposed sensing-aided V2V beam tracking solution requires data collected in a real wireless environment. This paper uses the publicly available scenario $36$ of the DeepSense 6G dataset. This scenario includes two vehicles traveling close to each other in the same direction of traffic. The collected dataset includes diverse real-world vehicular scenarios such as passing, lane changes, $4$-way intersections, and stop signs. The data was collected continuously with a sampling rate of $10$ samples/s, resulting in an initial raw dataset of $\approx 31$k samples. This raw dataset was subsequently processed by synchronizing and aligning data across modalities and cleaned up by manual data filtering. The new dataset after post-processing is made up of $\approx 21$k samples. The goal of this work is to observe a sequence of $r=5$, consisting of the current and next four $360\degree$ image samples, and predict the optimal beam index for the sixth sample in the sequence. Therefore, to generate the final development dataset for the beam tracking task, the new dataset is further processed using a sliding window to generate a time-series dataset consisting of 5 input images, the normalized mmWave power vector of the first sample in the sequence, and the optimal beam index of the sixth sample. In order to train the proposed transmitter identification model, as described in Section~\ref{sec:proposed_soln}, we require the ground-truth bounding box center coordinates of the transmitter in the scene. For this, we manually annotate $600$ image samples, which are further split into training and testing samples with a 70/30 ratio. We adopted the K-fold validation method for the beam tracking task, randomly splitting the $21$k data sequences into $5$ folds, each consisting of $\approx 4.2$k sequences.

\begin{table}[!t]
	\caption{Design and Training Hyper-parameters}
	\centering
	\setlength{\tabcolsep}{5pt}
	\renewcommand{\arraystretch}{1.2}
	\begin{tabular}{@{}l|cc@{}}
		\toprule
		\toprule
		\textbf{Parameters}                     & \textbf{FC Neural Network}  & \textbf{LSTM network}        \\ \midrule \midrule
		\textbf{Batch Size}                     & 32                  & 32                  \\
		\textbf{Learning Rate}                  & $1 \times 10 ^{-2}$ & $1 \times 10 ^{-3}$ \\
		\textbf{No. Hidden layers}            & 2     &  2   \\
		\textbf{Nodes Per Hidden layer} & 256                 & 150                 \\
		\textbf{Weight Decay}                   & $1 \times 10 ^{-4}$ & $1 \times 10 ^{-4}$ \\
		\textbf{Dropout}                        & 0.0                 & 0.1                \\
		\textbf{Total Training Epochs}          & 100                  & 200                  \\ \bottomrule \bottomrule
	\end{tabular}
	\label{tab_nn_train_params}
\end{table}

\begin{table}[!t]
	\caption{Performance Evaluation of the Proposed Solution}
	\centering
	\setlength{\tabcolsep}{8pt}
	\renewcommand{\arraystretch}{1.5}
	\begin{tabular}{|l|ccc|}
		\hline
		\multirow{2}{*}{\textbf{Sub-Tasks}} & \multicolumn{3}{c|}{\textbf{Model   Performance}}                                             \\ \cline{2-4} 
		& \multicolumn{1}{c|}{\textbf{top-1}} & \multicolumn{1}{c|}{\textbf{top-5}} & \textbf{R2-Score} \\ \hline \hline
		Transmitter Identification          & \multicolumn{1}{c|}{99.19}         & \multicolumn{1}{c|}{-}              & 90.23            \\ \hline
		Beam Tracking   (K=1)               & \multicolumn{1}{c|}{44.68}         & \multicolumn{1}{c|}{84.69}         & -                 \\ \hline
		Beam Tracking   (K=2)               & \multicolumn{1}{c|}{46.10}         & \multicolumn{1}{c|}{83.01}         & -                 \\ \hline
		Beam Tracking   (K=3)               & \multicolumn{1}{c|}{43.05}         & \multicolumn{1}{c|}{85.39}         & -                 \\ \hline
		Beam Tracking   (K=4)               & \multicolumn{1}{c|}{46.13}         & \multicolumn{1}{c|}{85.56}         & -                 \\ \hline
		Beam Tracking   (K=5)               & \multicolumn{1}{c|}{44.68}         & \multicolumn{1}{c|}{84.69}         & -                 \\ \hline \hline
	\end{tabular}
	\label{fcn_perf}
\end{table}

\section{Experimental Setup}\label{sec:exp_set}
In this section, we present the neural network training parameters and the adopted evaluation metrics. 

\subsection{Network Training}
As described in Section~\ref{sec:proposed_soln}, we trained and tested two different neural networks in our proposed solution. In the transmitter identification stage, the mmWave receive power vectors are provided as input to the FCN model to predict the polar coordinates of the transmitter object, which is subsequently utilized to approximate the bounding box center coordinates of the transmitter. The FCN is trained using the mean square error loss function and the AdamW optimizer. In the beam prediction stage, the 2-layered LSTM model is trained using categorical cross-entropy loss function and Adam optimizer. The LSTM model takes the sequence of transmitter coordinates from the transmitter tracking algorithm and predicts the future optimal beam. PyTorch and TensorFlow deep learning frameworks were utilized for the training, validation, and testing of the FCN and LSTM models, respectively. All the simulations were performed on a single NVIDIA Quadro 6000 GPU. The detailed design and training hyper-parameters are presented in Table~\ref{tab_nn_train_params}.



\begin{figure}[t]
	\centering
	\includegraphics[width=1.0\linewidth]{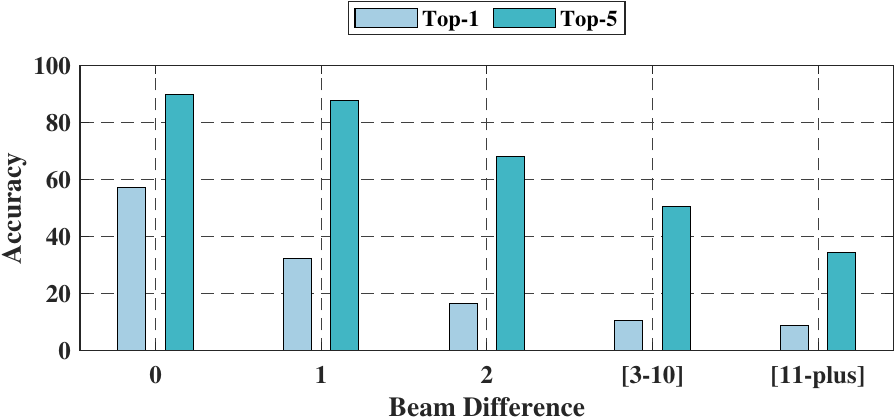}
	\caption{This figure shows the performance of the proposed V2V beam tracking solution versus the beam difference between the first and the fifth sample in the each sequence.  }
	\label{bfig}
\end{figure}

\subsection{Evaluation Metrics}
To evaluate the effectiveness of our proposed solution, we use top-k accuracy as the primary metric. This metric measures the percentage of test samples for which the optimal ground-truth beam is among the top-k predicted beams. To evaluate the effectiveness of our proposed solution, we use top-k accuracy as the primary metric. The top-k accuracy measures the percentage of test samples for which the optimal ground-truth beam is included in the top-k predicted beams. In particular, we consider the top-$1$ and top-$5$ accuracies to provide a comprehensive evaluation of our proposed solution.

\section{Performance Evaluation} \label{sec:perf_eval}
Given the experimental setup described in Section~\ref{sec:exp_set}, in this section, we study the beam tracking performance of the proposed vision-aided V2V beam tracking solution. We first evaluate the performance of the proposed transmitter identification solution, followed by an in-depth analysis of the overall beam tracking performance.

\textbf{Can visual and wireless data be utilized for transmitter identification?} As the proposed solution is a multi-stage algorithm, we first analyze the performance of the transmitter identification model based on evaluation metrics such as the top-1 accuracy and R-squared score between the predicted and ground truth transmitter coordinates, as shown in Table~\ref{fcn_perf}. The top-1 accuracy is the probability of successfully selecting the true transmitter coordinates based on the output of the FCN model and the nearest distance-based algorithm. It is important to note here that the results for the transmitter identification task are based on $30\%$ of the manually annotated $630$ samples discussed in Section~\ref{sec:proposed_soln}. From Table~\ref{fcn_perf}, it is observed that the proposed solution achieves $99.19\%$ transmitter identification accuracy on this small manually annotated dataset. The high accuracy of the proposed approach highlights that the sensing-aided solution can help identify the transmitter in the $360\degree$ images captured by the receiver vehicle with high fidelity.

\textbf{Can visual and wireless data be utilized for beam tracking?} As presented in Section~\ref{sec:dataset}, the development dataset of $21$k data sequences is randomly split into $5$ folds, each consisting of $\approx 4.2$k sequences. This was done to ensure that there is no data leakage that might impact the overall beam-tracking performance of the proposed solution. In Table~\ref{fcn_perf}, we present the top-$1$ and top-$5$ beam tracking performance for the $5$ different folds. It is observed that the proposed beam tracking solution achieves an average top-$1$ accuracy of $\approx 45\%$ and a top-$5$ accuracy of $\approx 85\%$. Considering the top-$5$ beam prediction accuracy, it can be inferred that the proposed sensing-aided beam tracking solution can predict the optimal beam index with notable accuracy while significantly reducing the beam training overhead. One promising way to satisfy the reliability and latency requirements of real-world deployment can be to augment lightweight beam training along with the proposed sensing-aided solution. Therefore, instead of performing an exhaustive beam search over the entire beam codebook  (which has a size of $4Q=256$ for this dataset, considering the four phased arrays), we can obtain the optimal beam by performing beam training over only the top-$5$ predicted beams. Next, we present an in-depth analysis of the proposed beam-tracking solution.

\begin{figure}[t]
	\centering
	\includegraphics[width=1.0\linewidth]{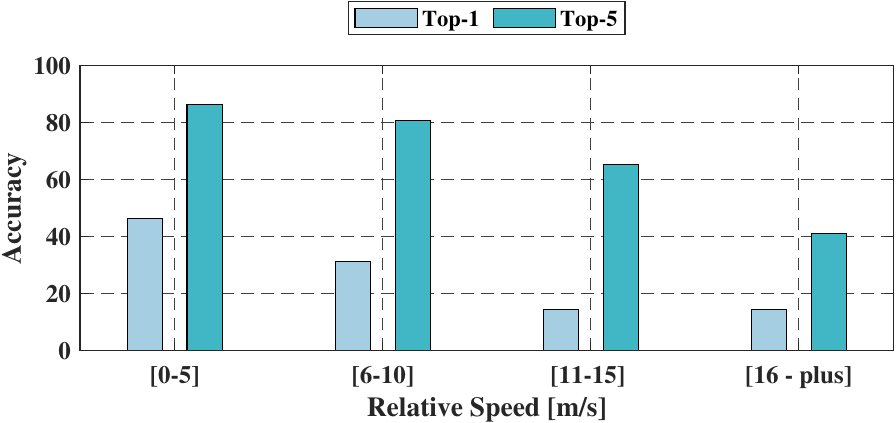}
	\caption{This figure shows the performance of the proposed V2V beam tracking solution versus the relative speed of the transmitter and receiver vehicle.}
	\label{sfig}
\end{figure}

\begin{figure}[t]
	\centering
	\includegraphics[width=1.0\linewidth]{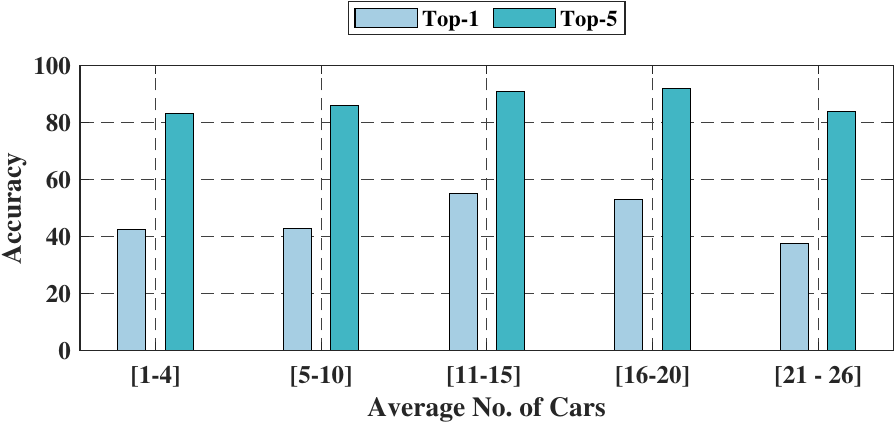}
	\caption{This figure shows the performance of the proposed V2V beam tracking solution versus the average number of objects detected in each sample.}
	\label{afig}
\end{figure}

\textbf{Impact of beam difference on the beam tracking performance:} The objective of this work is to observe a sequence of $r=5$ current and previous $360\degree$ image samples and predict the optimal beam index for the fifth sample in the sequence. In this section, we investigate the beam tracking performance of the proposed solution versus the beam difference between the first and the fifth samples in the sequence. It is important to note that the beam difference between the first and the fifth sample in a sequence captures the relative displacement between the transmitter and the receiver vehicle in the real world. Therefore, a significant beam difference within a sequence (for example, beam difference of $\ge 3$) can translate to vehicles overtaking each other, lane changes, and turns. As such, these sequences are generally far more challenging than those with smaller beam differences. We present the performance of the proposed sensing-aided V2V beam tracking solution in Fig.~\ref{bfig} for varying beam differences within a sequence. It is observed that the performance of the model decreases as the beam difference between the first sample and the fifth sample increases. Nonetheless, as shown in this figure, even for beam differences of approximately $10$, our solution achieves a top-5 beam prediction accuracy of over $50\%$. This further highlights the following: (i) The proposed solution can predict the optimal beam index with high fidelity for even the most challenging real-world scenarios, and (ii) augmenting our sensing-aided wireless communication systems with a lightweight beam training method is crucial to achieving optimal performance in practical wireless environments.

\textbf{What is the impact of the relative speed between the transmitter and receiver vehicle?} As presented in Section~\ref{sec:dataset}, scenario $36$ of the DeepSense 6G dataset adopted in this work includes two vehicles (transmitter and receiver vehicle) traveling near each other in the same direction. Our goal is first to identify the transmitter in the scene and then track the transmitter over time to predict the optimal beam index. Given the temporal nature of the problem statement, vehicular speed plays a critical role in determining the efficacy of any proposed solutions. Further, given the dynamic nature of the V2V communication dataset, the vehicles travel at different speeds, and their relative speed affects the beam tracking performance. Therefore, we study the impact of the relative speed between the transmitter and receiver on the beam tracking accuracy. Similar to the sequences with significant beam differences, higher relative velocity translates to considerable relative displacement between the transmitter and receiver vehicle. It poses a considerable challenge to the beam tracking task. In Fig.~\ref{sfig}, we plot the beam tracking accuracy versus the relative speed between the transmitter and receiver vehicle. It is observed that the increase in relative velocity impacts the overall beam tracking performance. However, as shown in Fig.~\ref{sfig}, even when the relative speed is more than $16$ m/s or $\approx 57$ Kph, the model can predict the top-5 beams with more than $40\%$ accuracy, highlighting the efficacy of the proposed solution.

\textbf{What is the impact of the number of objects detected in the visual scene?} The first two stages of the proposed solution include (i) object detection and (ii) transmitter identification. We adopt a state-of-the-art object detection model (YOLOv7) in the object detection stage to detect and identify the relevant objects in the wireless environment. However, one key challenge here is the missed objects, i.e., the objects that the model does not detect. For instance, if the transmitter is not detected in the first stage, the downstream beam tracking performance will also be adversely impacted. Further, as the number of objects in the visual scene increases, the chances of missed object detection also increase. The increase in the number of objects also impacts the next stage, i.e., transmitter identification. As described in Section~\ref{sec:proposed_soln}, the proposed transmitter identification solution utilizes the mmWave receive power vector to estimate the approximate center coordinates of the transmitter in the scene. The object in the visual scene with the shortest distance to this predicted value is then picked as the transmitting candidate. Now, as the number of objects increases, the chances of wrongly identifying a non-transmitting candidate as the transmitter also increase. All these highlight that the changes in the number of objects in the visual scene can impact the beam tracking performance. In Fig.~\ref{afig}, we plot the beam tracking accuracy versus the average number of objects detected in a sequence. We observe that the proposed beam tracking solution achieves a stable performance irrespective of the average number of relevant objects in the $360\degree$ camera scene for the samples within a sequence. However, it is important to note that the dataset does not have an evenly distributed number of objects in the samples. The average number of cars in the range $1-4$ has more samples, approximately $45\%$, followed by the range $5-10$ with about $25\%$. The range $21-26$ has the least number of samples, which is less than $1\%$ of the whole dataset.

\section{Conclusion}\label{sec:conc}
This paper explores the potential of leveraging visual sensory data for beam tracking in a mmWave V2V communication system. We formulate the vision-aided V2V beam tracking problem and develop an efficient machine learning-based solution to predict the optimal beam indices. Next, to evaluate the efficacy of the proposed solution, we adopt a real-world multi-modal V2V communication scenario from the DeepSense 6G dataset. The evaluation results demonstrate that the proposed vision-aided solution can learn to identify the transmitter in the visual scene, track the transmitter over a sequence of RGB images, and predict the optimal beam index with high fidelity. In particular, the proposed solution achieves a top-$1$ accuracy of $\approx 45\%$ and a top-$5$ accuracy of $\approx 85\%$ in predicting the optimal beams. These results highlight the potential of leveraging the visual sensors to enable highly-mobile mmWave V2V communication systems.

\balance

\end{document}